\RequirePackage{fix-cm}
\documentclass[smallextended]{svjour3}       
\smartqed  
\usepackage{graphicx}
\newcommand{\ligand}{\textit{ligand}}
\newcommand{\ligands}{\textit{ligands}}
\newcommand{\pocket}{\textit{pocket}}
\newcommand{\fragments}{\textit{fragments}}
\newcommand{\fragment}{\textit{fragment}}
\newcommand{\geodock}{\textit{GeoDock}}
\usepackage[ruled,linesnumbered]{algorithm2e} 
\usepackage{subfig}
\usepackage{dirtree}
\usepackage{forest}
\usepackage{prettyref}
\usepackage{listings}
\usepackage{subfig}
\usepackage[]{comment}

\newrefformat{eqn}{Equation~\ref{#1}}
\newrefformat{fig}{Figure~\ref{#1}}
\newrefformat{tab}{Table~\ref{#1}}
\newrefformat{sec}{Section~\ref{#1}}
\newrefformat{lst}{Listing~\ref{#1}}
\newrefformat{alg}{Algorithm~\ref{#1}}

\begin{document}

\title{
Exploiting OpenMP \& OpenACC to Accelerate a Molecular Docking Mini-App in Heterogeneous HPC Nodes
}

\titlerunning{Exploiting OpenMP \& OpenACC to Accelerate a Molecular Docking Mini-App}        

\author{Emanuele Vitali \and 
        Davide Gadioli \and 
        Gianluca Palermo \and
        Andrea Beccari \and 
        Carlo Cavazzoni \and
        Cristina Silvano 
}


\institute{Emanuele Vitali \and 
        Davide Gadioli \and 
        Gianluca Palermo \and Cristina Silvano \at
              Dipartimento di Elettronica, Infomazione e Bioingegneria, Politecnico di Milano, Italy.\\
              \email{name.surname@polimi.it}           
           \and
           Andrea Beccari \at
              Dompe' Farmaceutici SpA Research Center, L'Aquila, Italy.\\
              \email{andrea.beccari@dompe.com}
           \and
            Carlo Cavazzoni
            \at Cineca
             Supercomputing Innovation and Applicatioon Department, CINECA, Bologna, Italy.\\
             \email{c.cavazzoni@cineca.it}
            }

\date{Received: date / Accepted: date}

\maketitle

\begin{abstract}

In drug discovery, molecular docking is the task in charge of estimating the position of a molecule when interacting with the docking site. This task is usually used to perform screening of a large library of molecules, in the early phase of the process.
Given the amount of candidate molecules and the complexity of the application, this task is usually performed using High-Performance Computing (HPC) platforms. In modern HPC systems, heterogeneous platforms provide a better throughput with respect to homogeneous platforms.

In this work, we ported and optimized a molecular docking application to a heterogeneous system, with one or more GPU accelerators, leveraging a hybrid OpenMP and OpenACC approach.
We prove that our approach has a better exploitation of the node compared to pure CPU/GPU data splitting approaches, reaching a throughput improvement up to 36\% while considering the same computing node.

\keywords{Molecular Docking \and GPU \and CPU \and OpenACC \and OpenMP \and High Performance Computing }

\end{abstract}

\section{Introduction}
\label{intro}

The drug discovery process involves several tasks performed \textit{in-silico}, \textit{in-vitro}, and \textit{in-vivo}.
While the output of this process is a single valid solution, typically it starts considering a huge set of candidate molecules.
In the early stages of drug discovery, the focus is on finding a small set of candidate molecules, named \textit{ligands}, which have a strong interaction with the binding site of a target molecule, named \textit{pocket}.
These stages are performed \textit{in-silico} and they usually leverage molecular docking algorithms, which estimate the strength of the interaction between the target \pocket{} and the evaluated \textit{ligand}.

The strength of the interaction depends on the three-dimensional displacement of the \ligand{} when it interacts with the target \textit{pocket}.
Therefore, a docking algorithm must estimate the correct \ligand{} pose to test if the \ligand{} has to be discarded.
This is a complex task due to the high number of involved degrees of freedom.
Besides the six degrees of freedom for placing a rigid body in a three-dimensional space, it is possible to change the geometrical shape of a \ligand{} without altering its chemical properties.
Indeed, a subset of bonds between the \ligands{} atoms, named \textit{rotamers}, split the \ligand{} into two disjoint sets of atoms that are able to rotate independently along the bond axis, called \fragments{}.
Thus, each \textit{rotamer} introduces two degrees of freedom in the problem.
Given that a \ligand{} might have more than one hundred of \fragments{}, estimating the strength of the interaction between a \ligand{} and the target \pocket{} is a complex task.
If we consider that pharmaceutical companies would like to consider billions of \ligands{} to increase the probability to find the best candidates, we require a large computational power.
For this reason, pharmaceutical companies usually rely on High-Performance Computing to address the early phases of the drug discovery process.

The Top500 list \cite{top500} ranks HPC platforms according to their computational power in terms of throughput.
If we consider the most powerful platforms, as in November 2018, they use heterogeneous computation units.
Given that the performance of a platform is limited by the power consumed, energy efficiency is becoming a key factor in the HPC context.
Depending on the application algorithm, hardware accelerators such as GPU or Xeon-phi might significantly improve the application throughput with respect to general purposes CPUs, considering the same power consumption.
From the pharmaceutical company point of view, the benefits of an increment of the throughput are twofold.
On one hand, it might reduce the monetary cost of the drug discovery process.
On the other hand, it might increment the number of evaluated \ligands{} , therefore increasing the probability of finding a good candidate.

Given that each \ligand{} evaluation is independent from the others, the first stages of the drug discovery process are data parallel.
Moreover, the evaluation of each three-dimensional displacement of a \ligand{} is still independent of the others.
Given the high amount of independent computations, this process matches the GPUs computation paradigm.
In the typical heterogeneous programming model, the application is still executed in a general purpose CPU, named \textit{host}.
Only the hot-spots of the application are offloaded to accelerators, named \textit{devices}.
Typically, by using this approach application developers are facing two problems: (1) managing data transfers between host and device, and (2) implementing the hot-spots of the application using a different language for a different architecture.

In our previous work \cite{pbio}, we investigated the benefits and limitations of using the OpenACC \cite{openacc} language extension in a molecular docking application, to mitigate the second problem.
In this work, we implemented a hybrid version by using OpenMP \cite{openmp} and OpenACC to leverage the processing elements on a heterogeneous node.
In particular, given the limitation analyzed in the previous work, we aim at mapping each phase of the application on the most suitable processing element.
To summarize, the contributions of this work are the following:
\begin{itemize}
    \item We propose an OpenACC version of the geometric docking algorithm targeting GPU architectures;
    \item We propose a Hybrid CPU/GPU version of the geometric docking algorithm capable to fully exploit the node heterogeneity;
    \item We analyze the resource utilization of the different solutions to find the best configuration also in presence of multi-GPU nodes; 
    \item We discuss the obtained results, comparing them with the sequential CPU application and the GPU implementation of our previous work;
\end{itemize}

In Section \ref{sec:sota} we describe previous work in molecular docking. We focus on how they dealt with the complexity of the problem, to introduce and examine languages and techniques to accelerate and optimize computations.
Section \ref{sec:application} describes the original algorithm of the target molecular docking application, designed for general purpose CPU.
After this background, Section \ref{sec:gpu_application} describes how we used OpenACC and OpenMP to leverage the heterogeneity of the platform.
We evaluate the benefits and limitation of the enhanced application in Section \ref{sec:experimental_result}.
Finally, Section \ref{sec:conclusion} summarizes the overall findings and concludes the paper.

\section{State of the art}
\label{sec:sota}

Molecular docking is an important application domain in High-Performance Computing.
Due to the complexity of the problem, it is not possible to evaluate all the poses of a \ligand{} when it interacts with the target \textit{pocket}.
Therefore, all the algorithms proposed in literature aim at finding a \textit{good enough} solution within a reasonable amount of time.

A common approach for performing molecular docking employs stochastic algorithms to sample the space of possible \ligand{} poses.
For example, MolDock \cite{thomsen2006moldock} and AutoDock Vina \cite{trott2010autodock} leverage genetic algorithms to converge to an optimal solution.
While works such as Glide \cite{friesner2004glide} leverages a Monte Carlo algorithm.
Using a different approach, it is possible to use geometrical and chemical properties of the target \pocket{} and of the evaluated \ligand{} to drive the docking process following a heuristic.
For example, Dock 4.0 \cite{ewing2001dock}, FLEXX \cite{Kramer1999flexx}, and Surflex-Dock 2.1 \cite{jain2007surflex} belong to this category.
While stochastic algorithms converge to an optimal pose, the heuristic employed in deterministic approaches might fail to dock the \ligand{} in the target \textit{pocket}.
However, the reproducibility of the results might become a domain requirement since the later stages of a drug discovery process require a heavy monetary effort. 
In this work, we focus on LiGenDock \cite{beato2013use}, which is a deterministic docking application that might be used to perform two tasks.
On one hand, it aims at performing an accurate docking, by using chemical and physical information.
This task might be used to accurately estimate the three-dimensional pose of a \ligand{} to forward to later stages of the process.
On the other hand, it aims at performing a fast docking, using only geometrical information.
This task might be used to perform a fast virtual screening of a huge library of \textit{ligands}.

From the implementation point of view, previous works propose a different approach to leverage the available computation resources.
Frameworks such as LiGen \cite{beato2013use} and DOCK \cite{ewing2001dock}, support the MPI paradigm to enable multi-node scaling.
Other frameworks, such as Glide\cite{friesner2004glide}, MolDock\cite{thomsen2006moldock} or Autodock Vina\cite{trott2010autodock}, prefer to split the data across different nodes.
Therefore, each instance of the application executes in a different node, processing a fraction of the ligand database.
A post-processing phase is then required to unify and to evaluate the results.

From the single node perspective, DOCK\cite{ewing2001dock} and Autodock Vina\cite{trott2010autodock} have been designed to leverage the computational power of a homogeneous node. 
Differently, Glide\cite{friesner2004glide} and MolDock\cite{thomsen2006moldock} also address heterogeneous nodes. 

In this work, we use as baseline an application \cite{beato2013use} that is already able to scale horizontally.
We focus on vertical scaling addressing heterogeneous nodes with at least one GPU.
In this direction, two are the main approaches for offloading computation to GPUs: using dedicated languages and directive-based programming models.

In the first case, it is possible to write the source code of the application, using specific languages such as CUDA \cite{Nickolls:2008:SPP:1365490.1365500} or OpenCL \cite{Stone:2010:OPP:622179.1803953}.
Usually, they separate the source code executed by the CPU (host side), from the code executed by the accelerator (device side).
Therefore, application developers need (1) to write the offloaded kernel considering the device memory model and parallelization scheme.
Moreover, they need (2) to explicitly handle the data transfer between host and device; and they need to (3) write the boilerplate code for the initialization.
However, with this approach application developers have the finest control on the application source code.

The second approach leverages directive-based languages, like OpenACC~\cite{openacc} and OpenMP~\cite{openmp}. 
In these languages, the application developer uses compiler directives to annotate the source code.
The toolchain transforms and compiles the offloaded kernels, and generates the code to transfer the data between host and device.
Moreover, it automatically generates the initialization code. %
The benefit of this approach is the ease of use.
The programmer writes the entire application with a single language independently from the actual target, i.e. CPU host or the accelerator. 
With this approach, a single source code can be executed on different hardware, thus enabling functional portability.
However, the application developer is still in charge of writing an algorithm suitable for the device memory model and parallelization scheme. 
Despite the multi-platform approach of those languages, the kernels must be tuned according to the target platform, since performance portability is still an open problem \cite{portability}. 
Indeed, one of the conclusions of our previous work \cite{pbio} was that the code modified and optimized for the GPU was not efficient when compiled and run on the CPU as the original code. 

To improve computation efficiency, our intention in this paper is to optimize the exploitation of all the computational resources available in an HPC node.
In modern systems, the HPC node typically includes several CPUs and GPUs.
The multi-GPU problem has been investigated in literature.
For example, the approaches proposed in \cite{homp,multipleomp} suggest to extend OpenMP to support multiple accelerators in a seamless way. 
OpenACC has runtime functions to support the utilization of multiple GPU, however, lacks GPU to GPU data transfer, in single node \cite{chapman2} and multinode \cite{markidis}.
A previous work in literature \cite{chapman} investigates a hybrid approach with OpenMP and OpenACC.
It proposes the usage of OpenMP to support a multi-GPU OpenACC application, assigning each GPU to an OpenMP thread.
In this context, each OpenMP thread performs the data transfer between the host and the target device, without performing any other computation.

In this paper, we extend our previous work \cite{pbio} by suggesting a new approach that offloads to the GPU the most compute-intensive kernels, using a hybrid approach of OpenMP and OpenACC.
In this way, we exploit multi-GPU nodes offloading to accelerators only the kernels that maximize the advantage of being run on the GPU.
Differently from previous approaches, we rely on CPU to compute kernels that are not suitable for GPUs.
Moreover, to maximize the utilization of the resources, we dedicate more than one OpenMP thread for each GPU.
In this way, we split the workload across the compute units, according to the characteristic of the compute units and of the algorithm.

\section{Target Application}
\label{sec:application}
 
This section introduces the target application by describing its algorithm and by providing timing analysis.
We omit implementation details that are not required to understand the proposed approach, for simplicity and clarity reasons.

\begin{lstlisting}[basicstyle=\scriptsize,float, caption=Pseudo-code of the original algorithm. ,label=lst:one, xleftmargin=2em]
load(pocket);
for (ligand : ligands)
{
    for (pose_id=0; pose_id<N,pose_id++){
        generate_starting_pose(pose_id, ligand);
        align_ligand (ligand, pocket)
        for(rep = 0; rep <num_repetitions; rep++){
            optimize_pose(ligand, pocket);
        }
    }
}
\end{lstlisting}

\subsection{Application Algorithm}

This work targets the \textit{LiGenDock} \cite{beato2013use} molecular docking application.
\textit{LiGenDock} uses a two-level approach for screening a large library of \ligands{}.
First, it performs molecular docking considering only geometrical features, to filter out the \ligands{} that are not able to fit in the target \pocket{}.
Then, it simulates the actual physical and chemical interaction of the most promising \ligands{} to find a more accurate estimation of their three-dimensional pose when interacting with the target \pocket{}.

In this paper, we focus on a \textit{LiGenDock} mini-app performing only geometrical transformations, named \geodock{}.
\geodock{} attempts to capture key computation kernels of the molecular docking application for the drug discovery implemented in LiGenDock and it exploits only geometrical features.
By developing \geodock{} in parallel with the new version of LiGenDock, application developers can work with system architects and domain-experts to evaluate alternative algorithms to better satisfy the end-user constraints or to better exploit the architectural features.
\geodock{} enables us a faster performance analysis and the optimization of the key kernels.

\prettyref{lst:one} shows the pseudocode of \geodock{} algorithm.
In the initial phase, it reads the target \pocket{} and the \ligand{} library (lines 1--2).
The algorithm that performs the molecular docking between a \ligand{}-\pocket{} pair is the body of the outermost loop (lines 4--10).
Given the size of the solution space, an exhaustive exploration is unfeasible.
Therefore, \geodock{} uses a greedy optimization process to explore the solution space (lines 6--9), guided by heuristics, with multiple restarts (lines 4--5). 
The docking algorithm is composed of two phases.
In the first phase \geodock{} considers a \ligand{} as a rigid body and it aims at finding the best orientation that fits the \pocket{} (line 6).
The best alignment is evaluated with a scoring function, that is called at every rotation.
The second phase addresses the internal degrees of freedom of the \ligand{}, and it aims at optimizing the displacement of its atoms (line 8).
In this phase, each \textit{fragment} of the \ligand{} is sequentially rotated to optimize the shape of the \ligand{} inside the target pocket. 
The optimization procedure is repeated to refine the optimization of the pose (lines 7--9).

\subsection{Application Profiling}

\begin{figure}
\centering 
\includegraphics[]{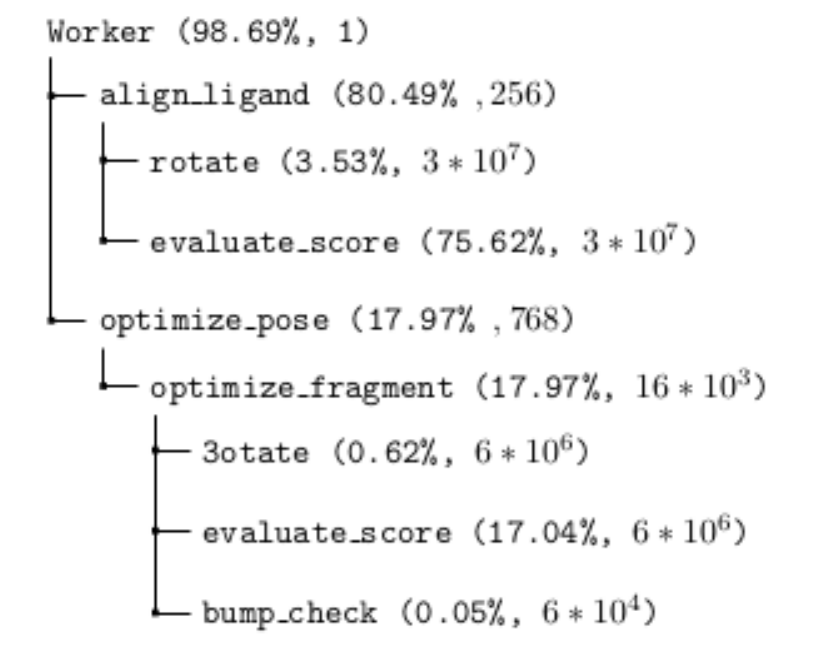}
\caption{The profiling of the application, in terms of percentage of execution time and number of visits.}
\label{fig:profile}
\end{figure}

To better understand the application complexity and to identify hotspots, we profile \geodock{} by using the Score-P \cite{scorep} framework.
\prettyref{fig:profile} shows the result of this analysis for the most significant functions.
In particular, it reports the percentage of time spent in the evaluated function and the number of times that the function is called.

From the results, we noticed how the scoring function is the bottleneck of the application.
This function is rather simple and it is already optimized for leveraging CPU architecture features, such as vector units.
In fact, the call to the function consumes less than 100$ns$.
The problem lies in the number of times that we need to call the function to dock a \ligand{} inside the \pocket{} ($10^7$).
Indeed, the scoring function evaluates ``how good'' is the current position of the \ligand{} with respect to the \pocket{}.
Therefore, we need to call the scoring function after every change of the \ligand{} shape.

From the profiling report, we might also notice how the functions that actually rotates the \ligand{} atoms (\texttt{Rotate}) or that tests whether a pose is valid (\texttt{BumpCheck}), have a negligible impact on the overall execution time, since they are also able to exploit the vector units of the CPU.
The two main kernels, i.e. the alignment kernel (\texttt{allign\_ligand}, line 6 in \prettyref{lst:one}) and the pose optimization kernel (\texttt{optimize\_pose}, line 8 in \prettyref{lst:one}), includes all these functions.

From the timing analysis, we might conclude how the performance of the application is not limited by a single complex function.
The bottleneck of the algorithm is due to the algorithm complexity for computing the \ligand{} pose, that needs to evaluate a high number of alternatives. 

Moreover, since the algorithm is greedy, we need to perform multiple restarts to reduce the probability of settling with a local minimum.
Therefore, since all of these operations are independent, it seems that the algorithm fits the parallel nature of the GPU.
The main problem is that we do not have a single kernel to offload to the GPU, but we need to consider the whole docking algorithm for the single \ligand{}.

\section{GPU Accelerated Versions}
\label{sec:gpu_application}
In this section, we describe the proposed approach to accelerate \geodock{} using GPUs.
First, we describe in Section \ref{sec:gpures} how we accelerate the whole algorithm with GPUs, relying on OpenACC.
Then, we describe in Section \ref{sec:hybrid} the analysis that lead us to implement the hybrid OpenMP/OpenACC solution.
In this second case, the idea is to allocate the workload on the heterogeneous resources according to different hardware capabilities, to improve the overall performance of the application.

\subsection{OpenACC Implementation}
\label{sec:gpures}

This paper extends our previous work \cite{pbio}, where we developed a pure GPU version of the algorithm, where both the main kernels (alignment and pose optimization) are offloaded to the GPU.
In particular, starting from the profiling analysis, we implemented a first version of the algorithm that aims at minimizing the data transfer while maintaining the application structure.

\begin{figure}
  \includegraphics[width=\textwidth]{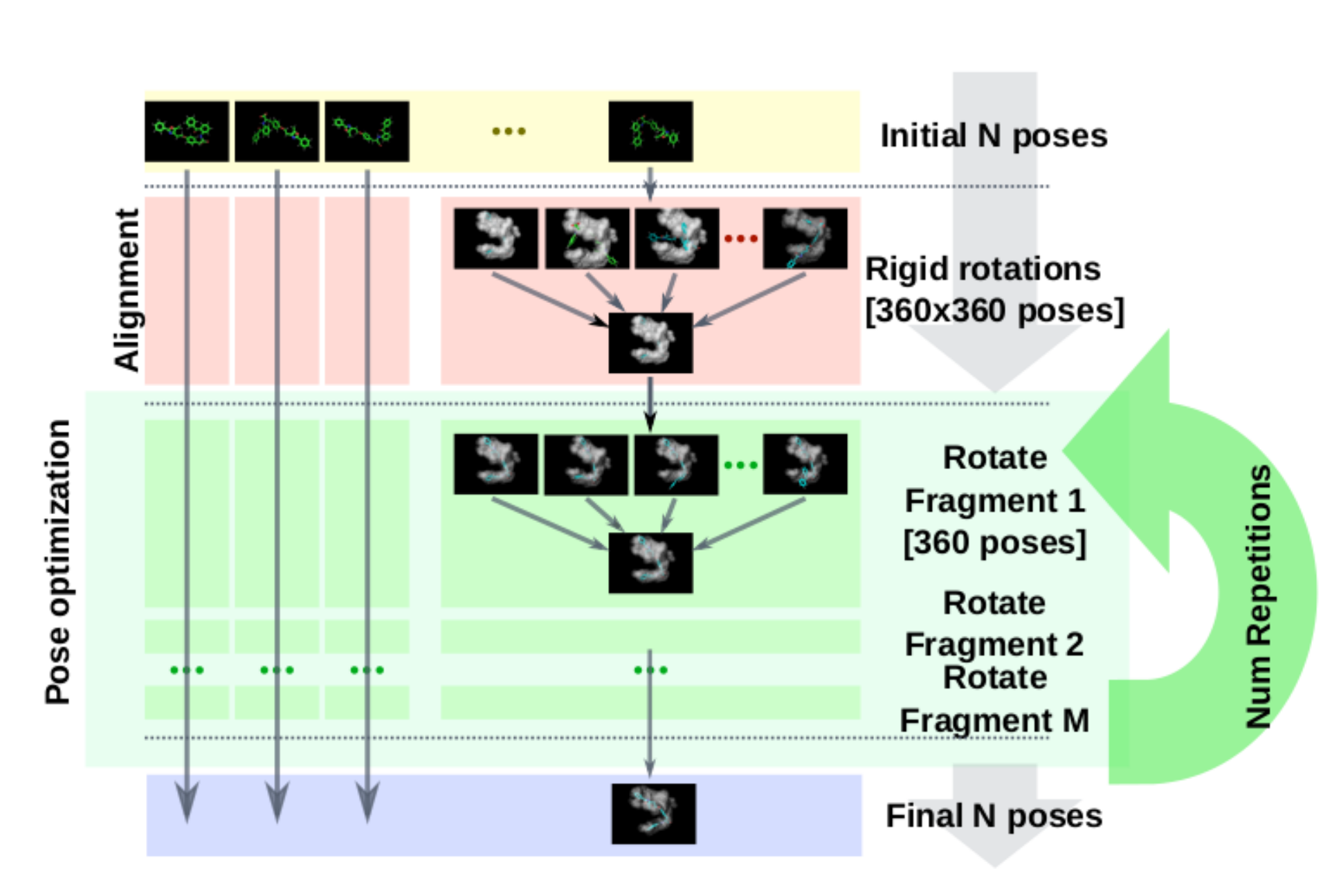}
\caption{Overview of the GPU implementation of the docking algorithm. Each box represents a computational part of the application that might be executed independently. The optimization phase must evaluate each \fragment{} sequentially and the whole procedure might be repeated to refine the final result. }
\label{fig:gpualg}       
\end{figure}

\prettyref{fig:gpualg} shows a graphical representation of the algorithm described in \prettyref{sec:application}, highlighting sections of the algorithm that are independent, by using different boxes.
We might consider each restart of the docking algorithm as a different initial pose.
Given that every initial pose might proceed independently, we have the first level of parallelism to map on the GPU.
Given an input \ligand{}, it is possible to generate and to dock the \ligand{} initial poses on the device side.
All the phases of the docking algorithm are performed in parallel, on different data, and we only extract the result at the end of the algorithm.
In this way, we transfer data only at the beginning and at the end of the docking algorithm (i.e. once in the lifetime of the \ligand{}).

From the implementation point of view, we use OpenACC to avoid rewriting the application source code in a different language.
OpenACC is able to operate with data structures that are resident on the GPU and usable across different kernels.
However, we need to introduce the following changes in the source code of the application to be compliant with the OpenACC standard \cite{OpenACCguide}.
First of all, the data structures that interact with the offloaded kernels must manage data transfers in the constructor and destructor.
In particular, the constructor allocates memory on the device side and it copies the initialized data into device memory.
The destructor must free both the device and the host memory.
Moreover, it is mandatory to mark each function called inside the offloaded region with the OpenACC \texttt{routine} directive.

Since our plan is to parallelize the computation over the initial poses, we need each pose in a different memory region to schedule all the iterations independently on the GPU.
The OpenACC language provides the \texttt{private} keyword to express this concept.
However, the system runtime available on our platform was not able to support this feature\footnote{The \geodock{} execution triggered an illegal accesses to the GPU memory when trying to transfer the private data structure.}.

Therefore, we had to manually duplicate the data.
With this modification, we were able to generate a correct binary.
However, the GPU application was slower than the baseline version running on the CPU.
We profiled \geodock{} to identify the cause of the slowdown since all the memory used by the application is resident on the GPU, therefore data transfers introduce a negligible overhead.
The result shows how the implementation is not able to exploit the parallelism.
In fact, the number of multiple restarts of the algorithm does not expose enough computation to the GPU.
In particular, using a single data structure to memorize the position of the \ligand{} atoms in the alignment function limits the amount of exposed parallelism.

To obtain an advantage from the use of the accelerator, we had to rework the source code to find (and expose) more parallel computation.
To achieve the desired result, we unified the rotation and scoring functions to avoid to store any temporary \ligand{} poses.
With this modification, we were able to expose more parallelism, since the rigid rotations are no more sequentially executed on a shared data structure.
After the computation, we schedule a reduction to retrieve the \ligand{} orientation with the best score.
Finally, we rotate the \ligand{} data structure accordingly, to forward to the optimization phase.
Therefore, we are able to replicate the parallelism depicted in \prettyref{fig:gpualg}.
We applied the same technique to further expose parallel computation also in the pose optimization phase.
However, the exposed parallelism is limited by two factors.
On one hand, we rotate a \fragment{} along with a 1-dimensional axis instead of a free rotation in a 3-dimensional space.
On the other hand, we must optimize each \fragment{} in sequence to save the \ligand{} consistency.
Therefore, we exploit again the pattern of parallel evaluation followed by a reduction, but the number of data is smaller: we need to perform a reduction at the end of every fragment evaluation.
This phase is repeated \texttt{num\_repetitions} times to improve the pose estimation.

\begin{lstlisting}[basicstyle=\scriptsize,float, caption=Pseudo-code of the GPU algorithm. ,label=lst:two, xleftmargin=2em]
load(pocket);
for (ligand : ligands)
{
    ligand_t ligand_arr[N];
    #pragma acc parallel loop
    for (pose_id = 0; pose_id < N, pose_id++){
        ligand_arr[pose_id] = ligand;
    }
    #pragma acc parallel loop gang
    for (pose_id = 0; pose_id < N, pose_id++){
        generate_starting_pose(ligand_arr[pose_id]);
        #pragma acc worker
        align_ligand(ligand_arr[pose_id], pocket);
        #pragma acc loop seq
        for(rep = 0; rep <num_repetitions; rep++){
            #pragma acc worker
            optimize_pose(ligand_arr[pose_id], pocket);
        }
    }
}
\end{lstlisting}

\prettyref{lst:two} describes the pseudocode of this GPU implementation.
With respect to the baseline algorithm, we replicated the original \ligand{} according to the number $N$ of multiple restarts of the algorithm (lines 4--8).
Once we initialize the memory on both the device and the host side, we evaluate each starting pose in the parallel region (lines 9--19) offloaded to the GPU.
It is possible to notice how in the pseudocode there are no pragmas for transferring data between host and device.
All the data transfers are managed with constructors and destructors of the data structures, according to the OpenACC standard.
To leverage all the levels of parallelism available in the GPU, we inserted different levels of parallelism in the code as well.
OpenACC offers three levels of parallelism: vector, worker, and gang.
Vector level parallelism is the SIMT (Single Instruction, Multiple Threads) level on GPU.
Gang level is the outer-most parallelism level, where all the elements are independent and the communication between gangs is forbidden.
Worker is an intermediate level used to organize the vectors inside a gang.
In particular, the vector and worker levels are the dimensions of a CUDA block, while the number of gangs is the CUDA grid.
Therefore, we split the initial poses at gang level, since all of them are independent (line 9--10).
We set all the internal functions (not shown in \prettyref{lst:two}) that change the position of the atoms at vector level.
The intermediate functions (i.e. \texttt{align\_ligand} and \texttt{optimize\_pose}) are set at worker level (lines 12--13 and 16--17).
The pose optimization loop (lines 15--18) is marked with a \texttt{loop seq} pragma.
This is mandatory to force the compiler to execute that loop in a sequential way. 

To evaluate the performance of this GPU version, we profiled the application and compared the results with the CPU baseline.
In the alignment phase, we obtained a good speedup (16x).
We noticed that the pose optimization was less suitable for GPU acceleration since too few operations per kernel were possible. 
Indeed, the sequentiality of the fragments and the control flow operations, inserted by the correctness checks, limit the reached speedup over the baseline CPU version.
The final speedup for this kernel was of only 2x.
Moreover, the profiling results show how the bottleneck of the application is changed.
With the GPU version, approximately the 70\% of the time is spent in the Optimize Pose kernel, while the Alignment takes less than 30\%.
This is a different result with respect to the profiling done on the baseline application on CPU.

Furthermore, we performed a parameter space exploration and tuning of the size of gangs, workers, and vectors to optimize the usage of the GPU, taking into account NVIDIA recommended best practices.
From experimental results, the best size configuration for each function is:
\begin{itemize}
    \item Alignment (\texttt{align\_ligand}): 8 workers and 128 vector length.
    \item Pose optimization (\texttt{optimize\_pose}): 64 workers with 1 as vector length.
\end{itemize}
This information confirmed us the hypothesis we made beforehand: the second kernel is less GPU-friendly than the first one.
More in-depth analyses of this version are described in previous work \cite{pbio}.

\subsection{Hybrid OpenMP/OpenACC Implementation}
\label{sec:hybrid}

\begin{figure}
\centering
  \includegraphics[width=0.8\textwidth]{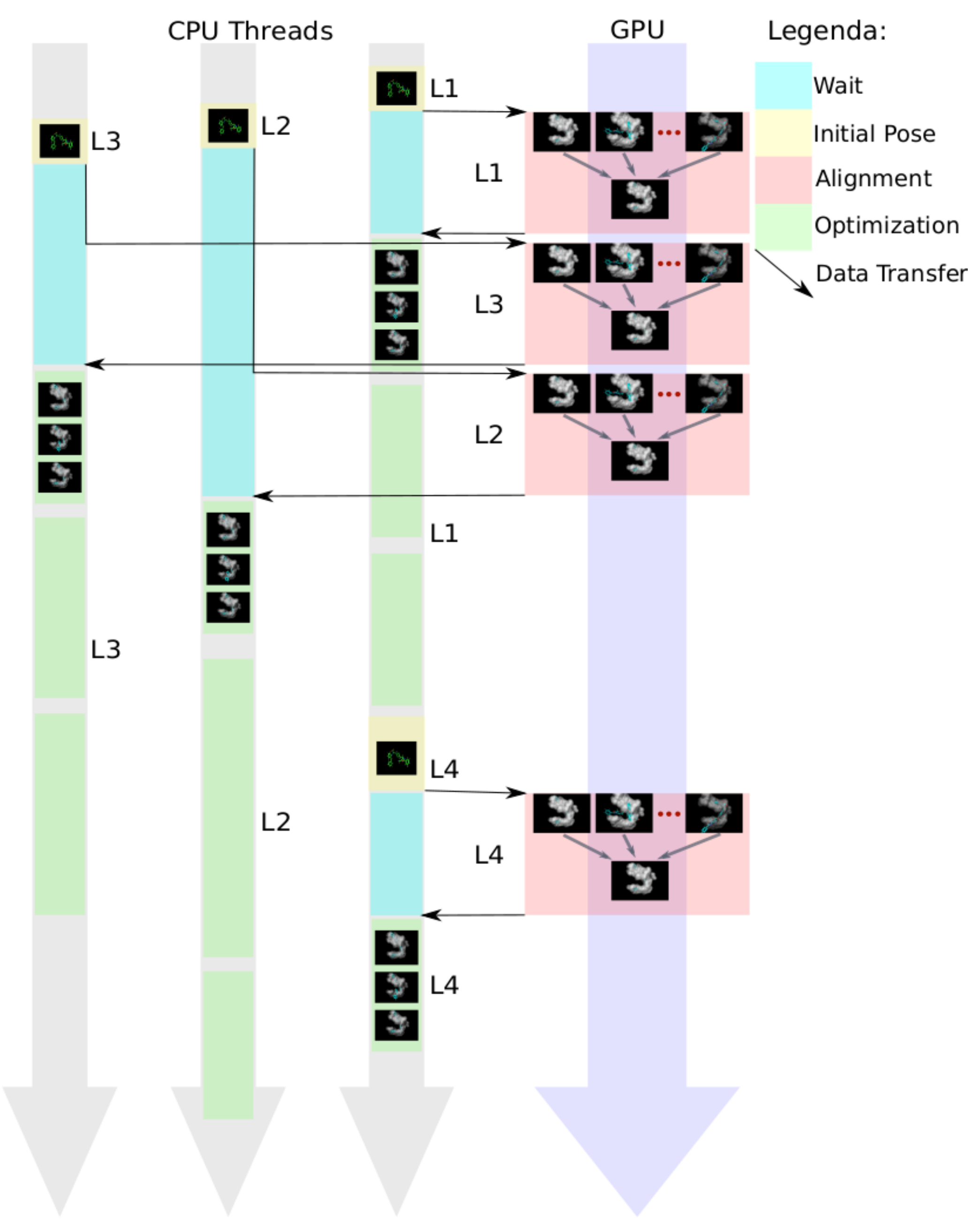}
\caption{Overview of the hybrid OpenMP/OpenACC implementation of the docking algorithm. The alignment phase of the \ligand{} is offloaded to GPU, while the optimization phase is perfomed in the CPU. Each OpenMP thread interact with a single GPU. The arrows identifies data transfer between host and device.}
\label{fig:2}       
\end{figure}

The \geodock{} application implemented in GPU, and described in the previous section, has two main limits.
On one hand, it is not able to use the available CPU cores to perform the computation.
On the other hand, not all the phases of the application are able to fully exploit the architectural features of the GPU.
As a consequence, the application is wasting or misusing a large fraction of the node computation capabilities.
Given that our target is to optimize the performance of \geodock{} on the full node, this section investigates the possibility to split the workload among CPU and GPU. 

From the profiling information of the GPU implementation, instead of simply partitioning the data among CPU processes and GPU processes, we modified the algorithm to bring the pose optimization phase back to the CPU. 
In particular, we would like to exploit the multicore architecture, enabling each CPU thread to evaluate one \ligand{}, and offloading only the alignment kernel to the GPU.
The basic idea, depicted in \prettyref{fig:2}, is to exploit the GPU for the kernel that benefits most of massive-parallel architecture, while mapping the other kernels to the CPU.
Each CPU thread takes care of a different \ligand{}, avoiding data movement among threads, therefore maximizing also the parallelism determined by the \ligand{} library. 
The only data movements are between CPUs and GPUs. 

\begin{lstlisting}[basicstyle=\scriptsize,float, caption=Pseudo-code of the hybrid algorithm.,label=lst:three, xleftmargin=2em]
load(pocket);
#pragma omp parallel
for (ligand : ligands)
{
    #pragma omp single nowait
    #pragma omp task
    {
        ligand_t ligand_arr[N];
        #pragma omp critical
        #pragma acc data
        {
            #pragma acc parallel loop
            for (pose_id = 0; pose_id < N, pose_id++){
                ligand_arr[pose_id] = ligand;
            }
            #pragma acc parallel loop gang
            for (pose_id = 0; pose_id < N, pose_id++){
                generate_starting_pose(ligand_arr[pose_id]);
                #pragma acc worker
                align_ligand(ligand_arr[pose_id], pocket);
            }
        }
        for (pose_id = 0; pose_id < N, pose_id++){
            for(rep = 0; rep <num_repetitions; rep++){
                optimize_pose(ligand_arr[pose_id], pocket);
            }
        }
    }
}
#pragma omp taskwait

\end{lstlisting}

In this \geodock{} implementation, we use OpenACC for the GPU kernel programming, while we exploit OpenMP for the CPU-level parallelism.
\prettyref{lst:three} shows the pseudocode of the algorithm.
The outermost loop that iterates over the \ligand{} library is parallelized using an OpenMP parallel region (line 2), where for every different \ligand{} we create a \texttt{single nowait} task (lines 5--6). 
\textit{Task} is a construct that was introduced in OpenMP $3$ and it is used to describe parallel jobs leaving the organization of the parallelism to the scheduler.
They are particularly effective for parallelizing irregular algorithms. 
The \texttt{single} keyword specify that an OpenMP region (in this case the \textit{task}) executes a single instance of the related region.
It is required to enforce that each \textit{task} is in charge of an iteration of the outermost loop.
The \texttt{nowait} keyword is used to notify the scheduler that the tasks are not synchronized, so it should not wait for the execution of the previous task to finish before scheduling the new one.

The execution time spent for docking a \ligand{} depends on several factors, such as the number of atoms, the number of fragments, and geometrical properties of both the target \pocket{} and \ligand{}.
Since these factors might drastically change between \ligands{} of the same library, we might consider our docking algorithm as an irregular application.
Therefore, the proposed implementation leverages the tasks construct to create a task for every \ligand{} to be docked.
As soon as an OpenMP thread becomes free, a pending task is assigned to it, until there is a task waiting to be executed.
Moreover, we use the tied task implementation to limit migration, restraining a task to be executed on the same thread that generated it.
Moreover, we bind each OpenMP thread to a physical core, by using the OpenMP environment variable \texttt{OMP\_PLACES=cores}.
In this way, we are able to associate a \ligand{} to one physical core, avoiding the extra movement of the data.

The GPU kernel is implemented inside an OpenMP \texttt{critical} region (line 9) to avoid race conditions. 
We also considered to use OpenACC features, such as asynchronous queues, however they performed worse than this implementation.
To wrap the kernel execution, we insert the \texttt{taskwait} pragma at the end of the parallel region (line 30).
In this way, we enforce thread synchronization only at the end of the library of \ligands{} to be docked.

The \geodock{} algorithm implementation is similar to the one described in \prettyref{sec:gpures}.
In particular, the data replication (lines 12--15) and the alignment phase (lines 16--21) are almost the same.
The only difference is in the data structure implementation, due to the limited support of C++ standard libraries from OpenACC.
For this reason, we manually managed data copies before and after the critical section used to offload the alignment to the GPU.
These changes are omitted in the application pseudocode.
However, we encountered a key issue in the memory management of the hybrid solution.
In the GPU version, we used CUDA unified memory to reduce the impact of data organization on the application developer. 
This feature enables addresses accessible from different types of architectures (normal CPU and CUDA GPU cores) hiding the complexity of the management from the programmer. 
If we use this implementation, the Unified Memory support for the Kepler architecture fails to properly allocate memory\footnote{When we enable the multi-threading with OpenMP, the CUDA managed memory fails. 
The manager tries to allocate the memory, from different threads, in the same area and returns a runtime error.}.
To solve this issue, we manually manage the memory allocation and transfers, by using OpenACC pragmas.
For this reason, we created a data region around the offloaded kernels (line 10).
The pose optimization kernel is no more decorated with pragmas (lines 23--26) because it is executed on the CPU, therefore we need to iterate over the aligned poses (line 23).

\textbf{Tuning considerations.} 
The hybrid approach requires a careful tuning to efficiently exploit the computing resources of the heterogeneous node. 
In fact, we can highlight two possible problems: GPU idle time and CPU thread waiting time. 
In the first case, the CPU threads are not able to provide enough data to fully exploit the GPU, leading to resource underutilization.
It is possible to notice this effect in \prettyref{fig:2} on the GPU side.
After the execution of the alignment phase of the \ligand{} L2, all the other CPU threads are still busy on the pose optimization phase.
Therefore, The GPU is in idle state until the alignment of the \ligand{} L4 is offloaded.
The second problem happens when there are too many CPU threads and the GPU is overloaded.
In this case, each CPU thread can have a long waiting time before accessing the GPU to offload the alignment kernel. 
In \prettyref{fig:2} we can notice an example of this problem at the beginning of the execution, where the \ligand{} L2 and the \ligand{} L3 are waiting for alignment of the \ligand{} L1 to end. 
For these two reasons, balancing the load between CPU and GPU is very important.
In the experimental results, we show how we tuned of the number of CPU threads to optimize the full node performance.

\begin{lstlisting}[basicstyle=\scriptsize,float, caption=Pseudo-code of the hybrid multi-GPU algorithm.  ,label=lst:four, xleftmargin=2em]
load(pocket);
omp_lock_t lock_array[N_GPUS];
#pragma omp parallel
for (ligand : ligands)
{
    #pragma omp single nowait
    #pragma omp task
    {
        ligand_t ligand_arr[N];
        omp_set_lock(lock_array[tid%N_GPUS]);
        #pragma acc set device_num(tid%N_GPUS)
        #pragma acc data
        {
            #pragma acc parallel loop
            for (pose_id = 0; pose_id < N, pose_id++){
                ligand_arr[pose_id] = ligand;
            }
            #pragma acc parallel loop gang
            for (pose_id = 0; pose_id < N, pose_id++){
                generate_starting_pose(ligand_arr[pose_id]);
                align_ligand(ligand_arr[pose_id], pocket);
            }
        }
        omp_unset_lock(lock_array[tid%N_GPUS]);
        for (pose_id = 0; pose_id < N, pose_id++){
            for(rep = 0; rep <num_repetitions; rep++){
                optimize_pose(ligand_arr[pose_id], pocket);
            }
        }
    }
    #pragma omp taskwait
}

\end{lstlisting}

\textbf{Multi-GPU.}
The considerations on the application tuning are even more important when we address multi-GPU nodes. 
From the implementation point of view, to distribute the workload across multiple devices, it is enough to provide different values to the \texttt{\#pragma acc set device\_num(...)}.
We used the thread number to decide on which GPU the thread will offload the kernel.

Moreover, we substituted the original critical section with an OpenMP mutex.
This gave us the possibility to exploit the parallelism in the kernel offloading having one kernel in each GPU.
The algorithm is reported in \prettyref{lst:four}.
In particular, we set the device using the related OpenACC pragma (line 11), after locking the mutex (line 10).
In this way, a set of threads is associated with a single GPU. 
As already mentioned, tasks are associated with a thread only when they start executing, and not at their creation.
This characteristic of tasks manages the load balancing.
\label{sec:lock}

\section{Experimental Result}
\label{sec:experimental_result}
We performed the experimental campaign using one GPU node of the GALILEO2 machine at CINECA\footnote{http://www.hpc.cineca.it/hardware/galileo-0}.
The target node is equipped with a 2x8-core Intel(R) Xeon(R) CPU E5-2630 v3 @ 2.40GHz CPU and two NVIDIA Tesla K80 GPU cards.
The operating system was CentOS 7.0, and we compiled the program using PGI Compiler 17.10 enabling the \emph{fastsse} flag to activate the vectorization on top of the \emph{O3} optimization level.

The data shown in this section are the results of several runs using different sets of \ligands{} and target \pocket{} to keep into consideration possible performance variability.
In particular, each experiment considers a single target protein \pocket{} with the goal of docking $1500$ \ligands{}.
This set of \ligands{} can be seen as a workload of a slave MPI process running on a single node in the context of a larger master-slave MPI application.
In the baseline CPU application, the master task sends a new batch of ligands as soon as a slave finishes its work.
The size of the \ligand{} set ($1500$) is big enough to ignore load balancing issues inside the node.
Therefore, we are able to measure the node performance in terms of average throughput.

To easily and fairly compare the different implementations we presented, we need to define the terminology that we use in the following sections.
We use the term CPU process $CPU^{proc}$ when we refer to an MPI process executing the baseline CPU version of the docking algorithm, described in \prettyref{sec:application}.
We use the term GPU process $GPU^{proc}$ when we refer to an MPI process executing the OpenACC version presented in \prettyref{sec:gpures}, which uses one CPU thread and one GPU.
Finally, we use the term hybrid process $HY^{proc}_{\#ompTh,\#GPUs}$ when we refer to an MPI process executing the OpenMP/OpenACC version described in \prettyref{sec:hybrid}.
In particular, $\#ompTh$ is the number of OpenMP threads, while $\#GPUs$ is the number of used GPUs.

\subsection{Single GPU}

\begin{figure}
\centering
  \includegraphics[width=0.8\textwidth]{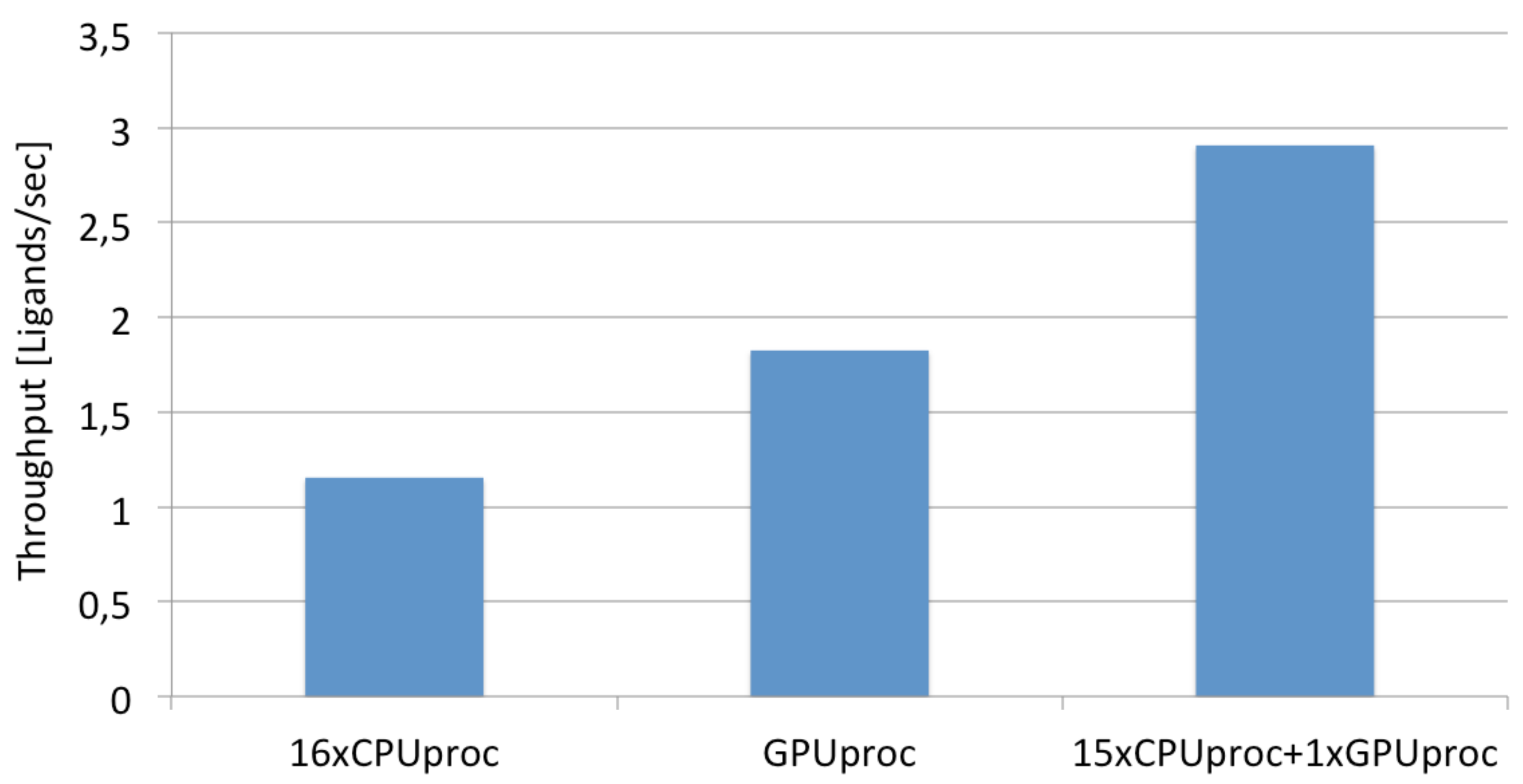}
\caption{Throughput of the \geodock{} application considering a single GPU, by varying its configuration.}
\label{fig:homogeneousperf}       
\end{figure}

This section analyzes the performance of the proposed hybrid implementation of \geodock{}, focusing on a single GPU case, to compare with previous implementations.
The first experiment aims at defining a baseline throughput, in terms of ligands per second, using the reference dataset.
\prettyref{fig:homogeneousperf} shows the throughput computed with different configurations of \geodock{} implementations: (i) by using all the CPUs ($16 \times CPU^{proc}$); (ii) by using the GPU ($1 \times GPU^{proc}$); (iii) by using the GPU and the remaining CPU cores ($15 \times CPU^{proc} + 1 \times GPU^{proc}$).
The number of $CPU^{proc}$ and the sum of $CPU^{proc}$ and $GPU^{proc}$ has been kept equal to the number of cores available in the node, because having more OpenMP threads than CPU cores does not increase the performance.
The results show how a single $GPU^{proc}$ has a speedup of 1.58x with respect the original application that exploits all the CPUs of the node ($16 \times CPU^{proc}$).
Therefore, if consider the configuration that uses the GPU and the CPU cores ($15 \times CPU^{proc} + 1 \times GPU^{proc}$), we are able to achieve a speed-up of 2.52x.
We use this third configuration as baseline for comparing the proposed hybrid approach. 

\begin{figure}
\centering
  \includegraphics[width=0.8\textwidth]{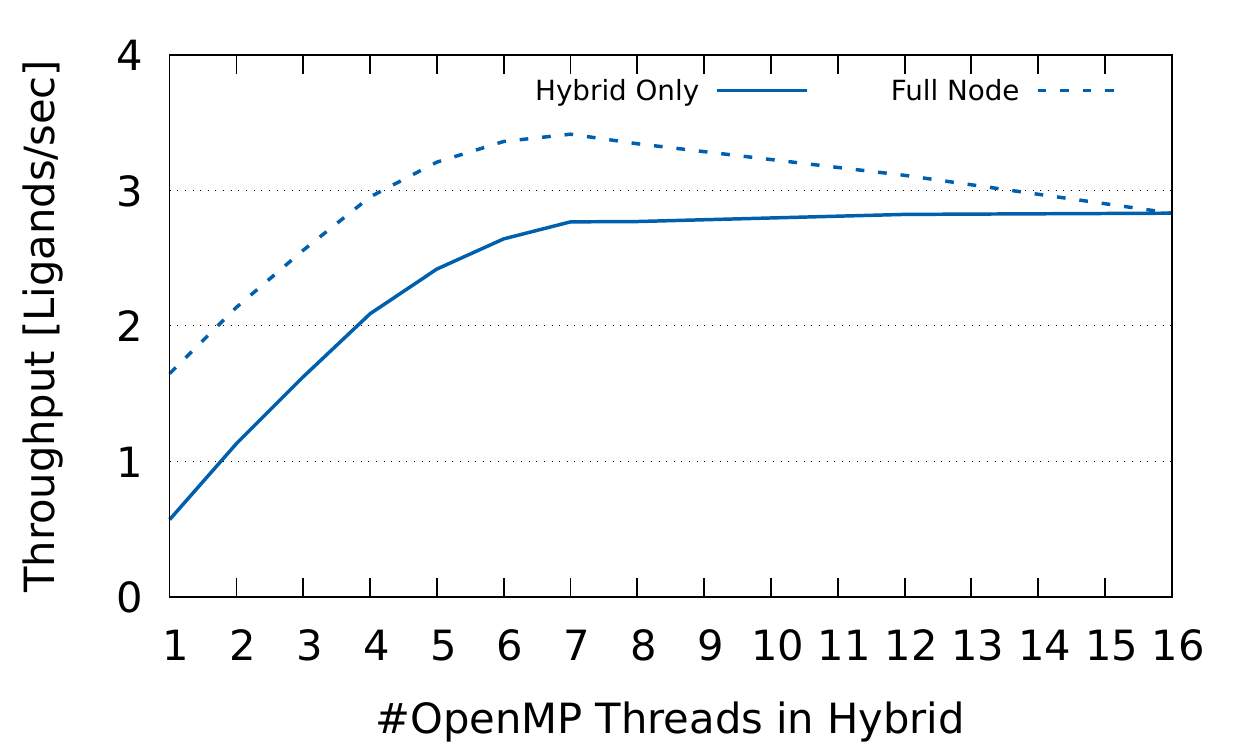}
\caption{Scaling analysis of the hybrid approach in terms of throughput, by changing the number of OpenMP threads.}
\label{fig:hybridsinglegpu}       
\end{figure}

The second experiment aims at analyzing the performance of the hybrid solution, by varying the number of OpenMP threads from $1$ to $16$.
\prettyref{fig:hybridsinglegpu} shows the experimental results.
In particular, the x-axis represents the number of the OpenMP threads used in the evaluated configuration  ($HY^{proc}_{n,1}$, with n=\{1,...,16\}).
The y-axis represents the reached throughput.
The solid line represent the throughput reached only by the $HY^{proc}_{x,1}$, while the dashed line represents the throughput when it is combined with a number of CPU processes sufficient to completely fill the node, i.e. $HY^{proc}_{n,1} + (16-n) \times CPU^{proc}$.

The results can be split into two different regions.
On the left side of the figure, we can see that increasing the number of OpenMP threads up to 7 the performance of $HY^{proc}_{n,1}$ almost linearly increases.
This is mainly due to the increment of GPU usage.
Few CPU threads are not able to fully exploit the GPU.
On the other hand, starting from 7 OpenMP threads the $Hy^{proc}_{n,1}$ performance reaches a saturation point since the GPU starts to be the bottleneck.
Indeed, at 7 OpenMP threads, the GPU is already fully used and with more CPU threads feeding it does not increment the throughput.
Similarly, the performance of $HY^{proc}_{n,1} + (16-n) \times CPU^{proc}$ reaches the maximum throughput when the saturation for the hybrid version happens (i.e. $HY^{proc}_{7,1} + 9 \times CPU^{proc}$). 
After this configuration, the performance of $HY^{proc}_{n,1} + (16-n) \times CPU^{proc}$ reduces while increasing the number of the OpenMP threads for the hybrid version.
If we use more CPU threads for the hybrid version, we are reducing the number of the CPU process we can exploit and thus their cumulative throughput contribution.
The result for the optimal configuration reports a speedup of 1.17x with respect to the $15 \times CPU^{proc} + 1 \times GPU^{proc}$ configuration, which was set as our baseline.

\subsection{Multi-GPUs}
This experiment analyses the performance of the hybrid approach according to the number of available GPUs in the target node.
The analysis is done by considering up to 4 GPUs which is the limit of our node (each K80 card includes 2 GPUs).

\begin{figure}
\centering
    \includegraphics[width=0.8\textwidth]{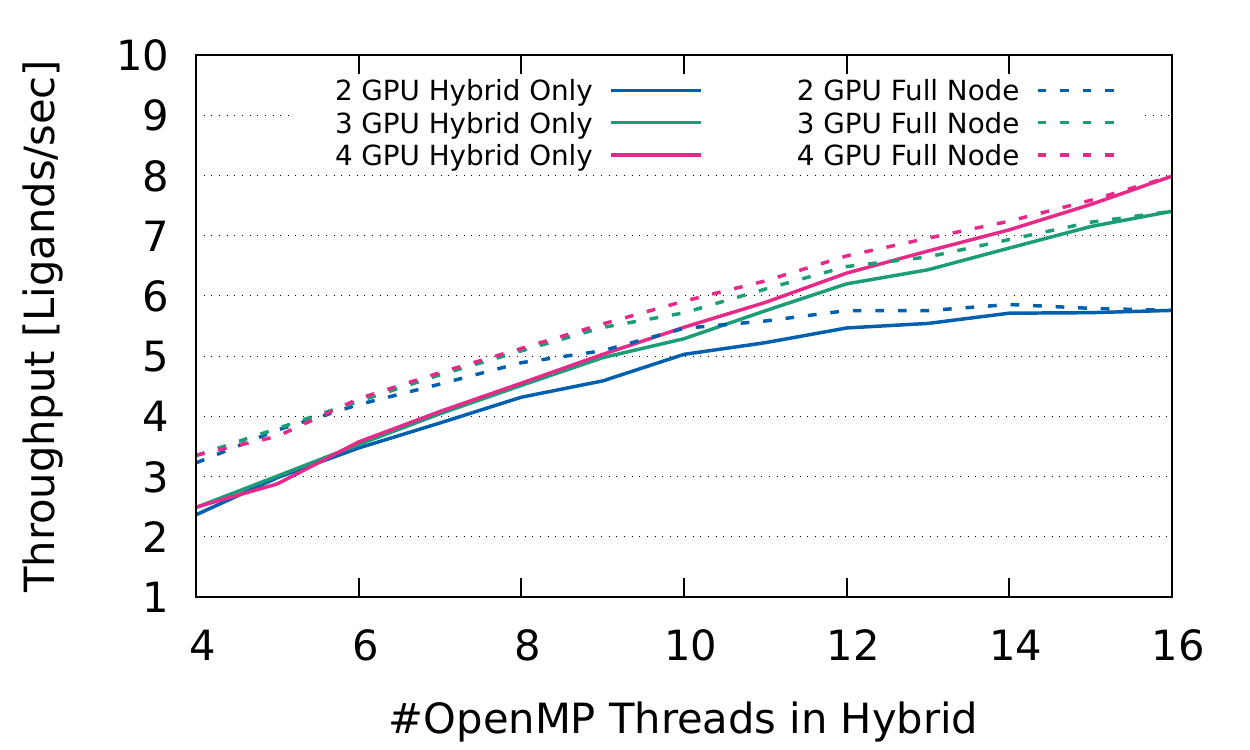}
\caption{Scaling analysis of the hybrid approach in terms of throughput, by changing the number of OpenMP threads and the number of GPUs.}
\label{fig:fullgpu}       
\end{figure}

\prettyref{fig:fullgpu} shows the application throughput of the hybrid process while varying the number of OpenMP threads (x-axis) and the number of GPUs.
In particular, solid lines represent \geodock{} configurations that uses only hybrid processes ($HY^{proc}_{n,k}$, with n=\{4, ..., 16\} and k=\{2, 3, 4\}).
While dashed lines represent the full node behaviour, where it uses $CPU^{proc}$ for the unused cores, i.e. the $HY^{proc}_{n,k} + (16-n) \times CPU^{proc}$ configuration.

If we focus on \geodock{} configurations that use only the hybrid approach with a node composed of two GPUs, experimental results shows how the application has an almost linear growth up to 8 cores (from 2.5 ligands per second to 4.5 ligands per second).
Then, the throughput gain slows down and it is almost negligible if we increase the number of OpenMP threads from 12 to 16 (the throughput ends at 5.8 ligands per second).
On the other hand, if we focus on \geodock{} configurations that use 3 and 4 GPUs, we have a steady growth in the application throughput over the entire range of OpenMP threads.
We might conclude that with the amount of OpenMP threads we considered, the GPUs are under-utilized.
In particular, the \geodock{} configurations that use 3 GPUs are slowing down a bit only in the last part of the plot, while in the \geodock{} configurations that use 4 GPUs the throughput grows almost linearly.

The dashed lines represent the usage of $CPU^{proc}$ for the spare cores. The results show a similar trend.
The main difference lies in the first part of the plot, where there are few OpenMP threads employed in the hybrid approach.
The maximum throughput for the full node is the maximum point of these lines.
We can notice from the picture that, according to the number of GPUs, the highest point of the functions is reached by the following configurations: $HY^{proc}_{14,2} + 2 \times CPU^{proc}$, $HY^{proc}_{16,3}$, and $HY^{proc}_{16,4}$.

\begin{figure}
\centering
  \includegraphics[width=0.9 \textwidth]{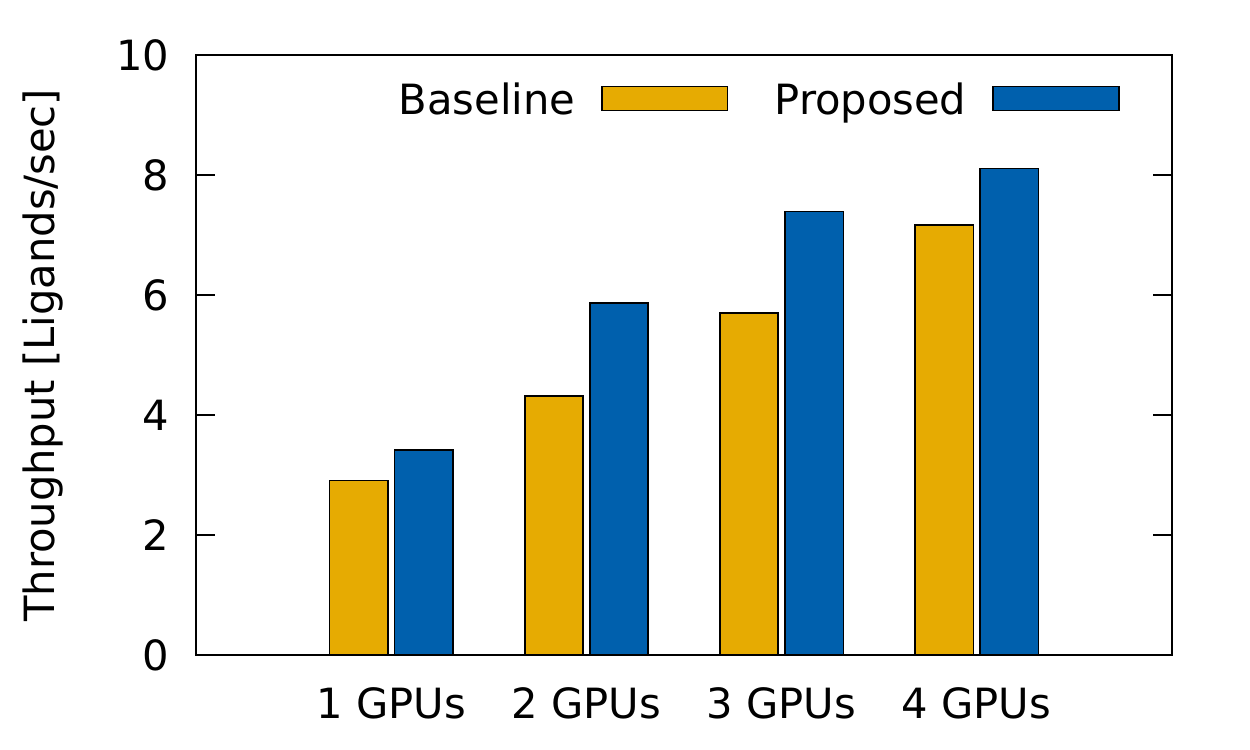}
\caption{Comparison of the baseline configuration of \geodock{} with the proposed hybrid approach, by changing the number of GPUs.}
\label{fig:final_comparison}
\end{figure}

To conclude our performance analysis, \prettyref{fig:final_comparison} shows the comparison of the best configurations exploiting the full heterogeneous node. As \emph{baseline} we used $k \times GPU^{proc} + (16-k) \times CPU^{proc}$ (where $k$ is the number of available GPUs) and as \emph{proposed} we select the best configuration obtained with the hybrid approach according to the previous analyses.
In all cases, by varying the number of available GPUs, the configurations including the hybrid version have a higher throughput.
This is due to the best exploitation of the GPUs only for the kernels where there is a higher speedup.
In particular, the performance improvement in the case of 1, 2, 3 and 4 GPUs is respectively of 17\%, 36\%, 30\%, and 13\%.
As mentioned before, in the case of 4 GPUs the performance speedup with respect to the baseline is lower because the 16 cores are not enough to fully exploit all the GPUs.

\section{Conclusion}
\label{sec:conclusion}
In the drug discovery process, the virtual screening of a huge library of ligands is a crucial job.
The benefits of an improvement in the execution time of this task, are twofold.
On one side it might reduce the cost of the process. On the other side, it enables the end-user to increase the number of the evaluated \ligands{}, increasing the probability of finding a better solution.

In this paper, we focused on a geometric docking application,\geodock{}, optimized for the CPU architecture of an HPC platform, and we ported it to GPU architectures.
First, we leveraged OpenACC to port the most intensive kernels on the GPU. Then, we modified the application flow, by limiting the GPU usage only to the kernels with the largest speedup, thus splitting the workload among the heterogeneous resources. This has been done by implementing a hybrid solution combining OpenMP and OpenACC to organize the computations.
We performed an experimental campaign to evaluate the scaling of \geodock{} when using all the available resources of the node. The experimental results have shown how the usage of the hybrid OpenMP/OpenACC version of \geodock{} guarantees a better exploitation of the resources of the node reaching a performance improvement up to 36\%.

\begin{acknowledgements}
This work has been partially funded by the EU H2020-FET-HPC program under the project ANTAREX - AutoTuning and Adaptivity appRoach for Energy efficient eXascale HPC systems (grant number 671623)
\end{acknowledgements}

\bibliographystyle{spmpsci}      

\bibliography{bla}   

\end{document}